\documentclass[pra,aps,twocolumn,showpacs,superscriptaddress,amsmath,amssymb]{revtex4-1}
\usepackage{color}

\usepackage{amsmath,graphicx}

\begin{document}
\def\tr{\rm{Tr}}
\def\la{{\langle}}
\def\ra{{\rangle}}
\def\a{{\alpha}}
\def\e{\epsilon}
\def\q{\quad}
\def\w{\tilde{W}}
\def\t{\tilde{t}}
\def\a{\hat{A}}
\def\h{\hat{H}}
\def\E{\mathcal{E}}
\def\p{\hat{P}}
\def\u{\hat{U}}

\title{Anomalous Zeno effect for sharply localised atomic states}
%
%
\author {D.  Sokolovski}
\affiliation{Departmento de Qu\'imica-F\'isica, Universidad del Pa\' is Vasco, UPV/EHU, Leioa, Spain}
\affiliation{IKERBASQUE, Basque Foundation for Science, E-48011 Bilbao, Spain}
\author {M. Pons}
\affiliation{Departmento de F\' isica Aplicada I, EUITMOP, Universidad del Pa\' is Vasco, UPV-EHU, Barakaldo, Spain}
\author {T. Kamalov}
\affiliation{Department of Physics, Moscow Open University, Moscow, Russian Federation}

\date{\today}
\begin{abstract}
We analyse the non-quadratic in time Zeno effect which arises when a few-atom state initially trapped between two high laser-induced barriers is briefly released to free evolution. We identify the Zeno time, analyse the energy distributions of those atoms which have escaped and those that remained inside the trap, and obtain a simple relation between the survival and non-escape probabilities. The relevant time scales are such that the effect would be observable for the atomic species used in current laser experiments.

\end{abstract}

%
%
\maketitle
%
%
%
%
%
%
%

The Zeno effect \cite{Z-1}-\cite{Z2} is one of the much studied basic phenomena predicted by the quantum theory. Its first experimental verification was reported in \cite{RAIZ1} where frequent observations were demonstrated to cause deviations from exponential decay law for sodium atoms trapped in an optical 'washboard' potential.
 Recent progress in laser technology has led to the creation of various types of atomic traps, among them a quasi-one-dimensional 'box' potential, in which atoms are confined between two laser-induced walls (endcaps) \cite{RAIZ2}. 
 For high laser intensities, the atomic wavefunctions are sharply localised inside the box  \cite{SH}, with only small exponential tails penetrating into the endcap laser beams.
Long- and medium-time evolution of few-atom states when one of the two endcap lasers is weakened or removed was studied in Refs. \cite{MUGA1,MUGA2}.
The subject of this Letter is the short time limit of this evolution.
As was demonstrated in \cite{RAIZ1}, ever more frequent interruption of the decay followed by a measurement of atomic population in the trap would affect the overall escape rate, 
which, as the Zeno effect sets in, will be completely suppressed.
One recalls that in a conventional Zeno effect, the probability to survive in the initial state $|\psi_0\ra$ 
decreases quadratically in time in the short time limit,
\begin{eqnarray}\label{1}
S(t)
=1-t^2/t^2_Z + O(t^2),\q
\end{eqnarray}
where the Zeno time $t_Z$ is determined by the energy spread of the initial state, ($\hbar=1$)
\begin{eqnarray}\label{2}
t_Z\equiv [\la\psi_0|\h|\psi_0\ra^2-\la\psi_0|\h^2|\psi_0\ra]^{-1/2}.
\end{eqnarray}
It follows that for a system subjected to frequent checks every $\tau$ seconds, the probability $W(t)$ to remain in $|\psi_0\ra$, is expected to exhibit an exponential decay
\begin{eqnarray}\label{3}
lim_{\tau \to 0}W(t)\approx \exp(-\gamma t),\q
  \gamma (\tau) \equiv \tau/\tau_Z^2,
\end{eqnarray}
which is stalled as $\tau\to 0$.
This analysis does not, however, apply to atoms trapped in a hard wall box potential.
In Ref.\cite{Schuss} the authors used the Feynman path integral approach to demonstrate
that the current, at the boundary of a sharply localised state subjected to free evolution, initially increases as $\sim \sqrt {t}$. With the probability to escape from the region of an initial localisation growing as $\sim t^{3/2}$ both Eq.(\ref{1}) and the definition of the Zeno time (\ref{3}) become meaningless.
It is the purpose of this Letter to investigate the anomalous (non-quadratic)  Zeno effect associated with this behaviour of the atomic current, as well as to suggest conditions for its experimental observation.

We start by considering a single atom of mass $M$ initially trapped between two nearly impenetrable laser-induced barriers of negligible width, placed at $x=0$ and $x=a$. The laser at $x=a$ is switched off and then restored again after a short time $t$, and we wish to evaluate the survival probability $S(t)$ to find the atom in its initial state in the trap.
It is natural to attempt to describe the system in terms of orthogonal left/right states spanning the trap interior for $0\le x\le a$ and the continuum at $x>a$, respectively,
\begin{eqnarray}\label{5}
\la x|\phi^L_n\ra =(2/a)^{1/2}\sin(\frac{n\pi x}{a})\theta_{[0a]}\q n=1,2,..\q\q\q\q\\ \nonumber
\la x|\phi^R_k\ra =(2/\pi)^{1/2}\sin[k(x-a)]\theta_{[a\infty]}\q k\ge 0,\q\q\q\q
\end{eqnarray}
Expanding  the atom's state $|\psi(t)\ra$ as
\begin{eqnarray}\label{6}
|\psi(t)\ra =\sum_n b_n(t)|\phi^L_n\ra +\int_0^{\infty} dk c_k(t)|\phi^R_k\ra, 
\end{eqnarray}
one easily  verifies that the Hamiltonian of the open trap,
\begin{eqnarray}\label{4b}
\h_0=(2/\pi)\int_0^{\infty}dk |\psi^0_k\ra\frac{k^2}{2M}\la\psi^0_k|,\q
\end{eqnarray}
where $\la x|\psi_k^0\ra=(2/\pi)^{1/2}\sin(kx)$, is diagonal in the left-right representation (\ref{5}). 
Thus, one has $\la \phi^L_n|\h|\phi^L_m\ra=E_n\delta_{mn}, \q\la \phi^R_k|\h|\phi^R_{k'}\ra=E(k)\delta(k-k'), 
\la \phi^L_n|\h|\phi^R_k\ra=\la \phi^R_k|\h|\phi^L_n\ra=0$, where
$E_n\equiv n^2\pi^2/2Ma^2$ and $ E(k)\equiv k^2/{2M}.$
The result is not particularly helpful, as it does not account for the atom's escape from an open trap and renders the Zeno time in Eq.(\ref{2}) infinite. Next we will show that the time evolution of the coefficients $b_n$ and $c_k$ is non-Hamiltonian, and their short time expansions contain not just integer, but also half-integer powers of $t$. For this purpose it is convenient to invoke the first- and last crossing time expansions \cite{FP1}-\cite{FP4} of the evolution operator $\u_0(t)\equiv \exp(-i\h_0 t)$,
\begin{eqnarray}\label{a1}
\u_0(t)=\u_L(t)+\u_R(t)+\q\q\q\q\q\q\q\q\q\q\q\q\q\\\ \nonumber
i\int_0^tdt_1 \u_0(t-t_1)\{[\h_0,\p_R]\u_L(t_1)+[\h_0,\p_L]\u_R(t_1)\} 
\end{eqnarray}
and
\begin{eqnarray}\label{a2}
\u_0(t)=\u_L(t)+\u_R(t)-\q\q\q\q\q\q\q\q\q\q\q\q\q\\\ \nonumber
i\int_0^tdt_1 \{\u_L(t-t_2)[\h_0,\p_R]+\u_R(t-t_2)[\h_0,\p_L]\}\u_0(t_2) \\ \nonumber
\end{eqnarray}
where $P_{L(R)}\equiv \int_{0(a)}^{a(\infty)}|x\ra\la x|$ are the projectors onto the interior (exterior) of the trap. The evolution operators $\u_{j}(t)=\p_j\exp(-i\p_j\h\p_jt)\p_j$, $j=L,R$ 
describe
the motion inside and outside the closed trap with an impenetrable narrow barrier at $x=a$, and $t_1$ and $t_2$ are the times at which the atom crosses the border of the trap at $x=a$ for the first and last time, respectively. With the help of Eq.(\ref{a1}), we see that an initial trap state  $|\phi_n^L\ra$ evolves into  $|\psi (t)\ra=\exp(-iE_n t)|\phi_n^L\ra +|\delta \psi(t)\ra,$ with $|\delta \psi(t)\ra$ given by the superposition of the waves emitted from the trap's border at $x=a$ for all $0\le t_1\le t,$ as shown in Fig. 1. 
\begin{figure}
\includegraphics[width=1\linewidth]{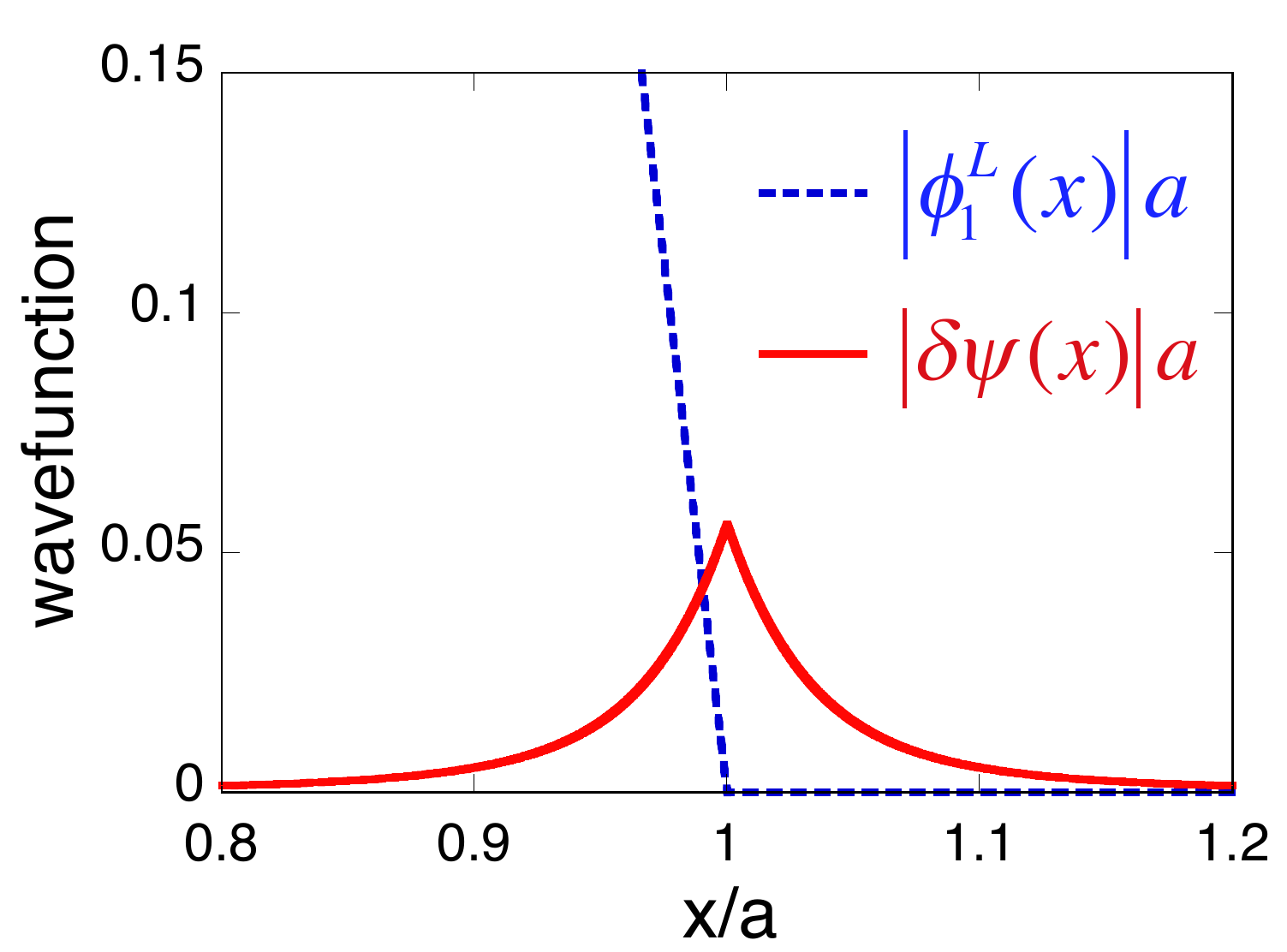}
  \caption{
\label{fcs}
(Color online)  $\delta \psi(x,t)$ obtained with Eq.(\ref{a1f}) propagating from the border of the trap at $x=a$, for $n=1$ and $t/t_0=0.001$(solid),
with $t_0\equiv Ma^2$.}
\end{figure} 
 Explicitly one has 
\begin{eqnarray}\label{a1f}
\la x|\delta \psi(t)\ra=(2m)^{-1}i\int_0^tdt_1 \la x|\u_0(t-t_1)|a\ra\times \\ \nonumber
\exp(-iE_nt_1)\phi_n^{L'}(a),
\end{eqnarray}
where $\phi_n^{L'}(a)\equiv \partial_x\la x|\phi_n^L\ra|_{x=a}$ and we have used the identity \cite{FP4} $\la f|[\h_0,\p_R]|g\ra=(2M)^{-1}[f'(a)g(a)-f(a)g'(a)]$. 
At short times, $E_nt<<1$, i.e.,
\begin{eqnarray}\label{a1a}
t<<2t_0/\pi^2n^2, \q t_0\equiv Ma^2
\end{eqnarray}
the matrix element $\la x|\u_0(t-t_1)|a\ra$ can be replaced by the free-particle propagator
$G_{free}(x-a,t)=\sqrt{M/2\pi it}\exp[iM(x-a)^2/2t],$ as it takes approximately a time $t_0$
for the wave emitted at $x=a$ to reach the laser beam at the origin \cite{FOOT-1}.
Combining Eqs.(\ref{a1}) and (\ref{a2}) we obtain the decomposition of the survival amplitude 
$A^{n\leftarrow n}(t) \equiv \la\psi^L_n|\exp(-i\h t|\psi^L_n\ra$ in terms of the first time the atom has left the trap, $t_1$, and the last time it has re-entered it, $t_2$, 
\cite{FP2}, \cite{FP5}
 \begin{eqnarray}\label{15a}\nonumber
A^{n\leftarrow n}(t)=-(2M)^{-2}|\phi_n^{L'}(a)|^2
\int_0^t dt_2\exp-iE_n(t-t_2)] \times \\ 
\int_0^{t_2}dt_1G_{free}(0,t_2-t_1)\exp(-iE_nt_1),\q\q\q\q\q\
\end{eqnarray}
which, in the short time limit (\ref{a1a}), reduces to
 \begin{eqnarray}\label{15}
A^{n\leftarrow n}(t)=\q\q\q\q\q\q\q\q\q\q\q\q\q\q\q\q\q\\\nonumber
1-iE_nt+
\frac{2^{1/2}n^2\pi^{3/2}}{3M^{3/2}a^3}\exp(-i\pi/4)t^{3/2}+O(t^2).\q\q 
\end{eqnarray}
Unlike in the conventional Zeno effect (\ref{1}),  the survival probability 
is not quadratic in time,
\begin{eqnarray}\label{16}
S(t)=1-t^{3/2}/t^{3/2}_Z + O(t^2),\q t_Z (n) \equiv \frac{3^{2/3}Ma^2}{2^{2/3}\pi n^{4/3}},\q\q
\end{eqnarray}
which makes it more difficult to preserve an atom it its initial state by frequent observations than 
 in the case of a conventional Zeno effect.
Indeed, subjected to frequent projection onto $|\phi^L_n\ra$ every $\tau$ seconds, the system will exhibit an exponential 
decay (\ref{3}) whose rate
\begin{eqnarray}\label{17}
  \gamma (\tau) \equiv \tau^{1/2}/\tau_Z^{3/2},
\end{eqnarray} 
would always exceed
that predicted by Eq.(\ref{3}) as $\tau \to 0$.

For the probabilities of transition to other states inside the trap Eq.(\ref{15a}) yields
\begin{eqnarray}\label{18b}
W_{m \leftarrow n}(t) \approx\frac{\pi n^2}{a}F(k_m,t),
\end{eqnarray} 
while the momentum distribution of the atoms which have escaped from the trap we have
\begin{eqnarray}\label{18a}
W_n(k,t) \approx  n^2 F(k,t)
\end{eqnarray} 
where
\begin{eqnarray}\label{18c}
F(k,t)\equiv\frac{2t}{a^3M^3k^2}
\times \q\q\q\q\q\q\q\q\q\q\q\q\\ \nonumber
|1-\frac{\sqrt{i\pi}}{2\sqrt{E(k)t}}\exp[-iE(k)t]Erf(\sqrt{-iE(k)t})|^2
\end{eqnarray} 
and $Erf(z)\equiv (2/\sqrt{\pi})\int_0^z\exp(-t^2)dt$ is the error function \cite{AST}.
The energy distributions $W_{m \leftarrow n}(t)$ and $W_n(k,t)$ are shown in Fig. 2.
\begin{figure}
\includegraphics[width=1\linewidth]{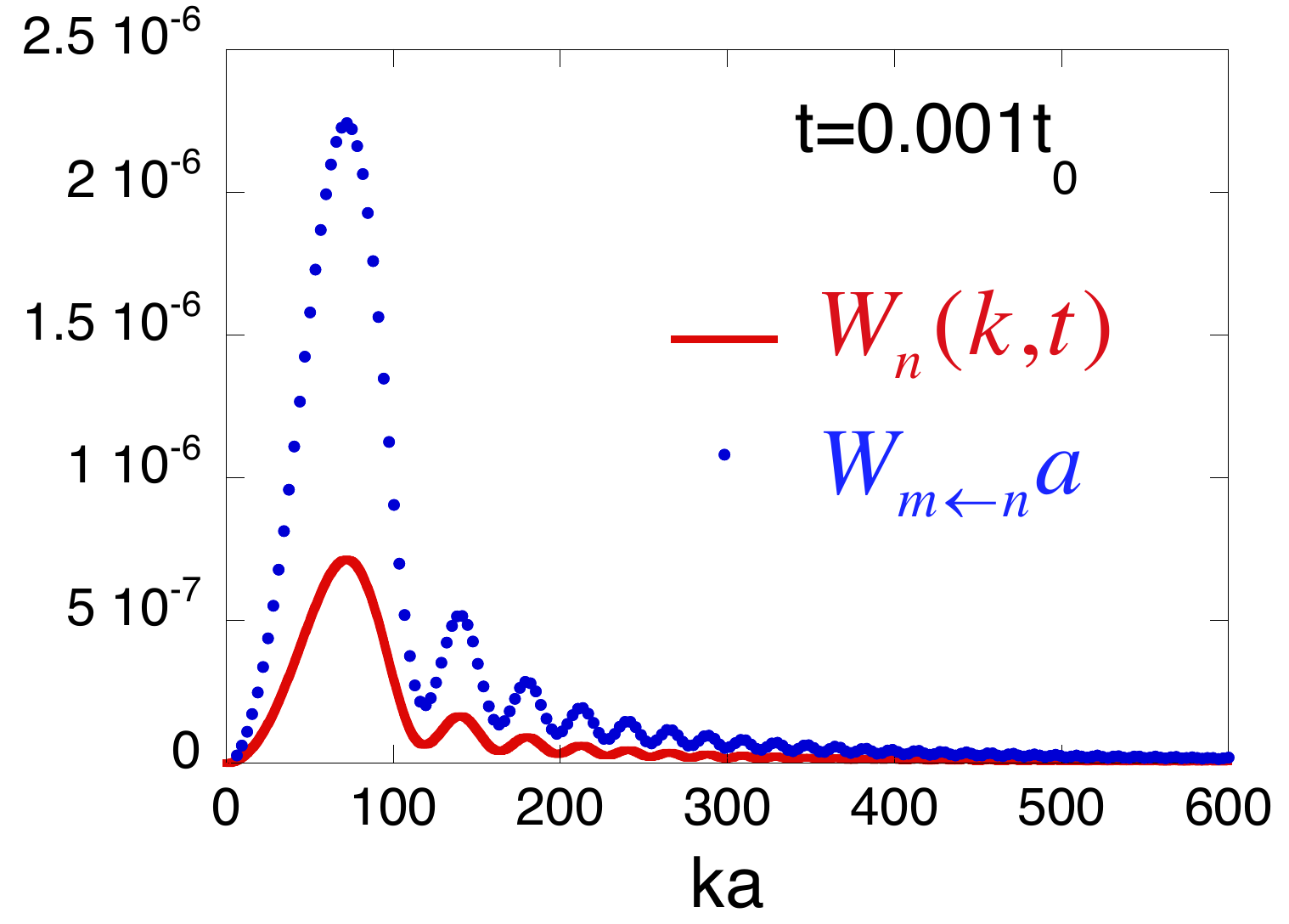}
  \caption{
\label{fcs}
(Color online) The momentum distribution of the escaped atoms $W_n(k,t)$ (solid) and the transmission probabilities
$W_{m \leftarrow n}a$ (circles) for $\delta \psi(x,t)$ shown in Fig. 1. }
\end{figure}

The non-escape probability $P_n(t)=1-\int_0^{\infty}W_n(k)dk$, i.e. the probability that the atom 
will remain inside the trap, although not necessarily in its initial state $|\phi^L_n\ra$, is more 
accessible to experimental observation than the survival probability $S(t)$.
(It is worth mentioning here that introduction of an additional hard wall at $x=b>a$ cannot affect the evolution of the system as long as $t<<Mb^2$. This allows us evaluate the number of escaped atoms from a discreet sum using the identity
 $\sum_{m=1}^\infty F(m\Delta k_m)\Delta k_m=\int_0^{\infty} F(k)dk$, $\Delta k_m\equiv \pi/b$ \cite{FOOT1}.)
Equations (\ref{18a}) and Eq.(\ref{18c}) show that the occupation of the low-energy states, with $Et<<1$, increases as $\sim t^3$, whereas for the high-energy states, $Et>>1$ we have 
$W_n(E,t)\sim t$. 
This complicates somewhat the evaluation of the short time behaviour of  $P_n$
directly from the integral $\int_0^{\infty}W_n(k)dk$ \cite{FOOT2}.
It can, however, be found using a variant of the optical theorem \cite{BAZ}. 
Indeed, equating the norm $\la\phi_n^L(t)+ \delta \psi(t)|\phi_n^L(t)+\delta \psi(t)\ra$ to unity yields $2Re \la\phi_n^L(t)|\delta \psi(t)\ra+\la \delta \psi(t)|\delta \psi(t)\ra=0$. Also, prior to expanding enough to reach the laser beam  at $x=0$, the state (\ref{a1f}) shown in Fig.1 is symmetric around the emission point, $\la a-x|\delta \psi(t)\ra = \la a+x|\delta \psi(t)\ra$, so that $P_n(t)=1-\la \delta \psi |\delta \psi(t)\ra/2$.  Furthermore, for the survival probability we have $S(t)=|\la\phi_n^L(t)|\phi_n^L(t)+\delta \psi(t)\ra|^2\approx 1+ 2Re\la\phi_n^L(t)|\delta \psi(t)\ra$, since the omitted term $|\la\phi_n^L(t)|\delta \psi(t)\ra|^2$ is negligible.
Combining the above, yields
\begin{eqnarray}\label{19a}
P_n(t)\approx [1+S_n(t)]/2 = 1-t^{3/2}/2t^{3/2}_Z + O(t^2).\\\nonumber
\end{eqnarray}

Using the one-particle amplitude (\ref{15}), one can evaluate short-time survival and non-escape probabilities for states containing several weakly interacting identical atoms  \cite{FERM1,FERM2}. For $N$ bosonic atoms occupying the same ground state, one easily finds the probabilities to survive in the state, $S^{(N)}(t)$, 
and that for all $N$ atoms to remain inside the trap, $P^{(N)}(t)$, given by Eqs.(\ref{16}) and Eq.(\ref{19a}), with $t_Z$ replaced by $t_Z^{(N)}=t_Z/N^{2/3}$, respectively. 

For fermionised bosons \cite{FERM} occupying the first $N$ states of the trap, one has $S^{(N)}(t)=|det_{n,k=1}^N[\la \phi_n(0)|\phi_k(t)\ra]|^2$, while the $N$-particle  non-escape probability is $P^{(N)}(t)=|det_{n,k=1}^N[\la \phi_n(t)|\p_L|\phi_k(t)\ra]|^2$, where the subscript $n$ ($k$) refers to the $n$ ($k$) -th one-particle state. For $t$ such that $E_nt<<1$, $n=1,2,...N$, expanding the determinants shows that $S^{(N)}(t)$ and $P^{(N)}(t)$  are given by Eq.(\ref{16}) and (\ref{19a}) with $t_Z^{(N)}=[\sum_{n=1}^Nt_Z^{-3/2}(n)]^{-2/3}$, respectively.

\begin{figure}
\includegraphics[width=1\linewidth]{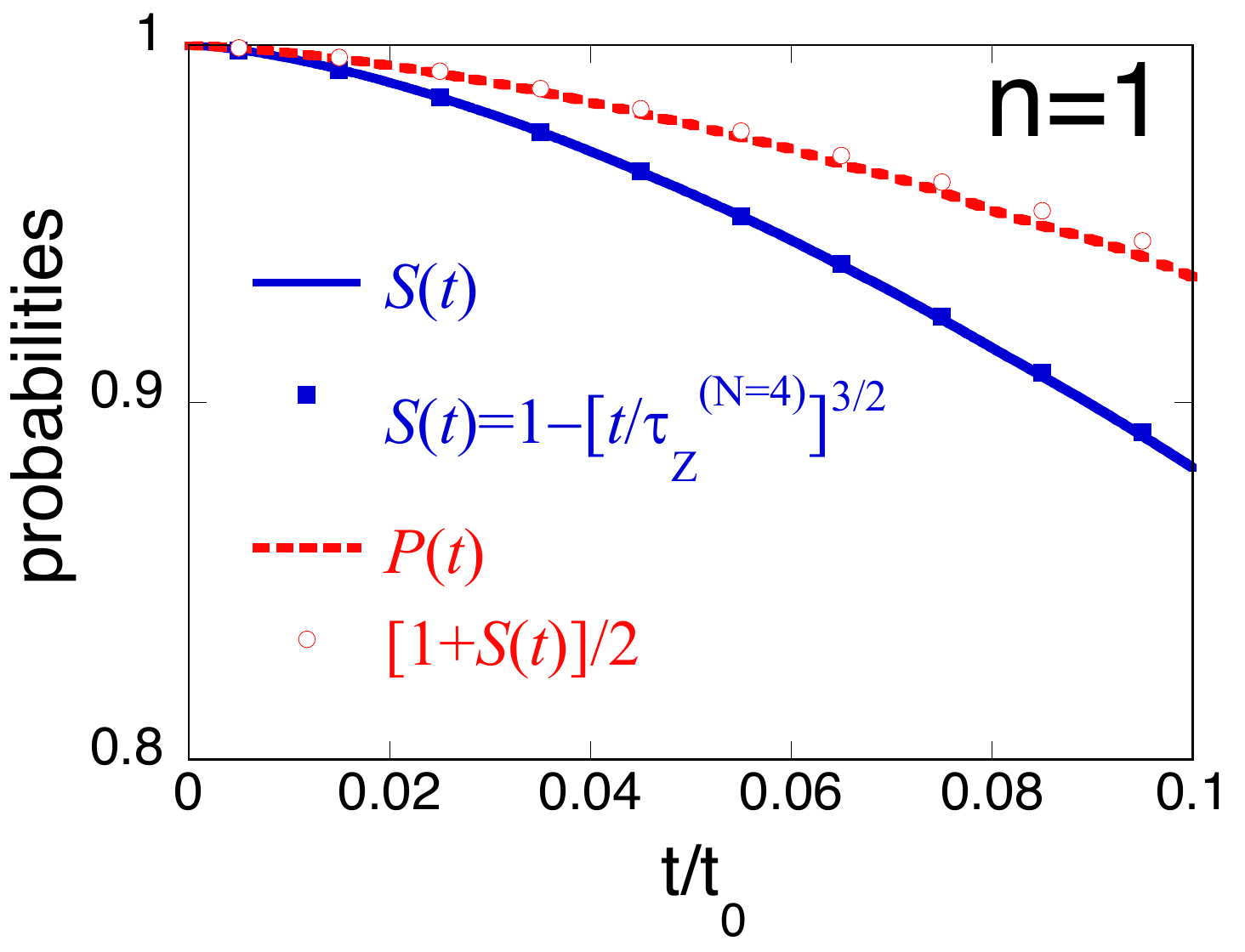}
  \caption{
\label{fcs}
(Color online) The survival probability $S(t)$ (solid)  and the non-escape probability  $P(t)$ (dashed) as functions of time for a single atom initially in its ground state $n=1$.
Also shown are $S(t)$ as given by Eq.(\ref{16}) (squares) and $P(t)$ in Eq.(\ref{19a}) (circles).
The value of $t_z/t_0$ is $0.418$.}
\end{figure} 

Although conveniently formulated for infinitely high and narrow barriers, our analysis 
remains valid for realistic potentials of a finite height, which allow penetration of a small tail of the initial wavefunction into sub-barrier region. Figs. 3 and 4 show the results obtained by numerically solving the time-dependent Schroedinger equation, for an atom initially in the ground state trapped between an infinite wall at $x=0$ and a potential step $V(x)=0$ for $x<a$ and $V_0t_0=(50\pi)^2$ for $x>a$.  Fig. 3 shows the dependence of the survival and non-escape probabilities vs. $t$ measured in the units of the time it takes the emitted wave $\delta \psi$ to reach the laser at $x=0$ (\ref{a1a}), 
$t/t_0$. Fig. 4 shows the same quantities for a system of four fermionised atoms occupying the 
four lowest levels in the trap. Since for $Rb$ and $Na$ atoms, and an initial trap of $a=80\mu m$, $t_0$ takes the values of $8.59 s$ and 
$2.32 s$, respectively, the non-quadratic behaviour and the anomalous Zeno effect would be observable, in the case of $N=4,$ shown in Fig. 4, at $t$   ($\tau$) of up to $0.15 s$ and $0.041 s$ respectively. both within 
capabilities of modern laser techniques.


In summary, in the orthogonal left-right representation, the short time evolution of a sharply localised wavefunction contains non-integer powers of $t$. This prevents one from defining the Zeno time in the usual way and leads to a non-quadratic Zeno effect and to an unusual behaviour of the energy distributions of the escaped and excited atoms.  As a result, 
the population of an open trap subjected to frequent checks would decay faster than for  a system exhibiting a conventional Zeno effect. This behaviour persists for the states which rapidly decay the confining potential barrier, and the relevant time scales are such that the effect should be observable for most of the atomic species used in current laser experiments. 

\acknowledgements

We acknowledge support of the Basque government (Grant No. IT-472-10), the European Regional Development Fund (ERDF) and the Ministry of Science and Innovation of Spain (Grant No. FIS2009-12773-C02-01 and No. FIS2008-01236). We are grateful to A. del Campo for useful discussions.

\begin{figure}
\includegraphics[width=1\linewidth]{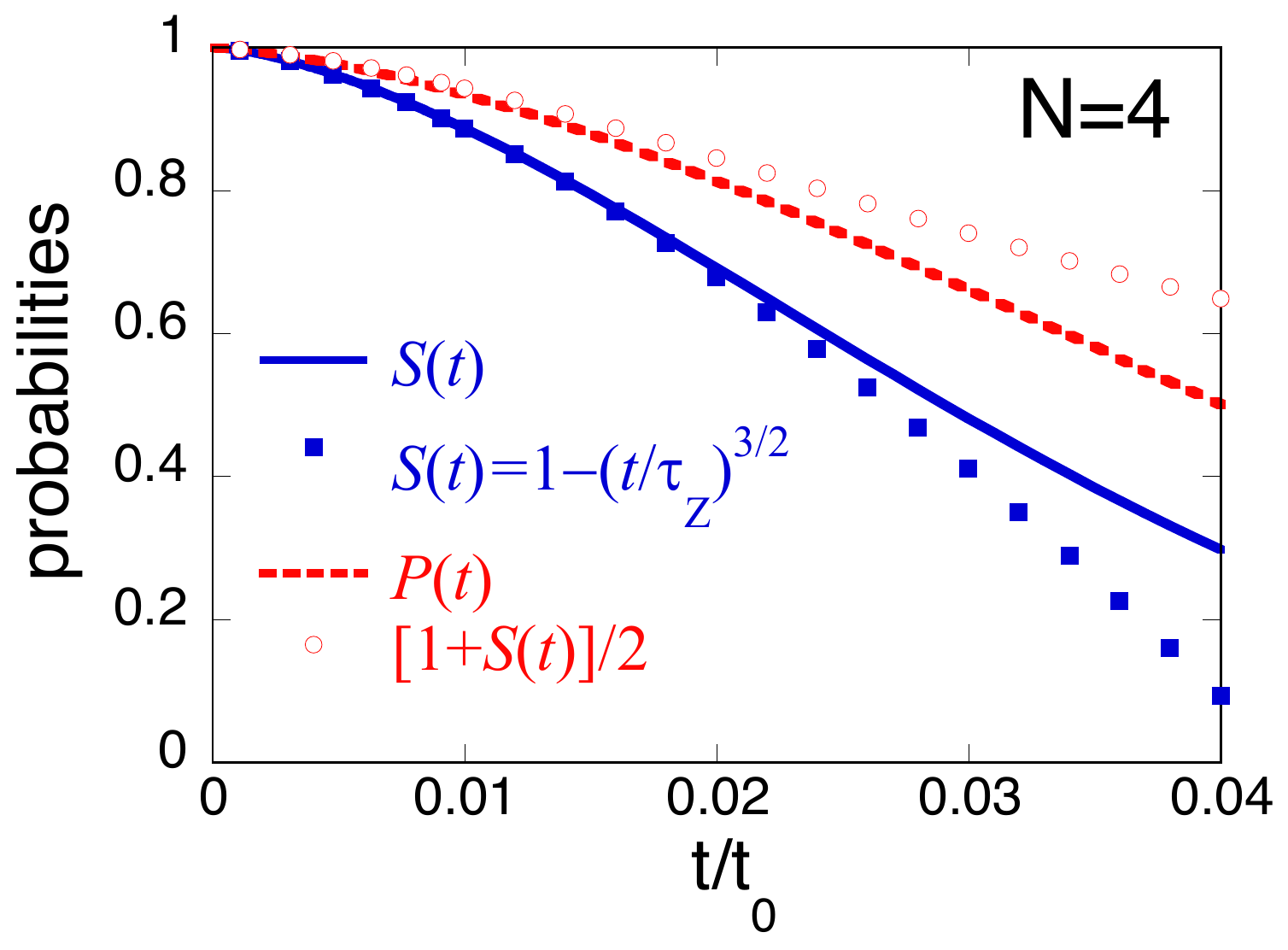}
  \caption{
\label{fcs}
(Color online) Same as Fig.3 but for four fermionised atoms occupying the four lowest states in the trap.
The value of $t_Z^{(N=4)}/t_0$ is $0.0427$.}
\end{figure}


\begin{thebibliography}{999}
\bibitem{FOOT-1} With $ \la x|\u_0(t-t_1)|a\ra$ replaced by $G_{free}(x-a,t)$, $\delta \psi(x,t)$ can be expressed in terms of the error function. Requiring $\delta \psi(0,t) << max[\psi_n^L(x)]$ yields 
 $t<<Ma2$.
\bibitem{FOOT1}
The $F(k)$ in Eq.(\ref{18c}) belongs to the class of functions sharing a curious property that for any $\Delta x \le \Delta x_{min}$, the Riemann sum $S(\Delta x)\equiv \sum_{n=1}^{\infty}F(n\Delta x)\Delta x$ is independent of $\Delta x$ and, therefore, equals exactly the integral $\int_0^{\infty}F(x)dx$. To put it differently, the shape of a function is such, that 
the area under the histogram $h(x)=F(n \Delta x)$ for $(n-1) \Delta x<x< n\Delta$, exactly matches the area under the curve $F(x)$ regardless of how the interval $[0,\infty)$ has been divided into equal parts. Such a function can be constructed, for example, by choosing $F(x)=|\int_0^a g(y)\sin(xy)dy|^2$ for which $\Delta x_{min}=\pi/a$. The property follows from the Parceval's theorem, which requires that the sum of the squares of the moduli of the coefficients of a Fourier series equal the integral $\int_0^a|g(x)|^2dx$, once the sine functions span the whole of the interval $[0,a]$ containing support of $g(y)$. For example, for $g(y)=0$ for $0<y<1$ and $0$ otherwise, one has
$F(x)=4x^{-2}\sin^4(x/2)$, $\Delta x_{min}=\pi$, and $S(\Delta x)=1$.
\bibitem{FOOT2} For comparison, in the conventional Zeno effect all transition probabilities are quadratic in time, and it is readily seen the probability to find the system in any subset of its Hilbert space containing the initial state must behave as $1-const\times t^2$.
\end{thebibliography}
\end{document}